\documentclass[
	journal
]{IEEEtran}

\IEEEoverridecommandlockouts % This is to make the \thanks{} command visible in conference mode
\makeatletter
\def\markboth#1#2{\def\leftmark{\@IEEEcompsoconly{\sffamily}\MakeUppercase{\protect#1}}%
\def\rightmark{\@IEEEcompsoconly{\sffamily}\MakeUppercase{\protect#2}}}
\makeatother

\renewcommand{\markboth}[1]{\renewcommand{\leftmark}{#1}\renewcommand{\rightmark}{#1}}
\usepackage[amssymb]{SIunits}
\usepackage[cmex10]{amsmath}
\usepackage{amssymb}
\usepackage{amsthm}
\usepackage[latin1]{inputenc}
\usepackage{cite}
\usepackage{url}
\usepackage{glossaries}
\usepackage{graphicx}
\usepackage[hang,small,bf]{caption}
\usepackage[subrefformat=parens,labelformat=parens]{subfig}
\usepackage{bm}
\usepackage[usenames]{color}

\newenvironment{DIFnomarkup}{}{}

\newacronym{AWGN}{AWGN}{additive white Gaussian noise} 
\newacronym{CSA}{CSA}{coded slotted ALOHA} 
\newacronym{SNR}{SNR}{signal-to-noise ratio} 
\newacronym{SINR}{SINR}{signal-to-interference-plus-noise ratio}
\newacronym{PLR}{PLR}{packet loss rate} 
\newacronym{UEP}{UEP}{unequal error protection} 
\newacronym{BS}{BS}{base station} 
\newacronym{LDPC}{LDPC}{low-density parity-check} 
\newacronym{VN}{VN}{variable node} 
\newacronym{CN}{CN}{check node} 
\newacronym{DE}{DE}{density evolution} 
\newacronym{MDS}{MDS}{maximum distance separable} 
\newacronym{BEC}{BEC}{binary erasure channel} 
\newacronym{PEC}{PEC}{packet erasure channel} 
\newacronym{DAMA}{DAMA}{demand assignment multiple access}

\newcommand{\figref}[1]{Fig.~\ref{#1}}

\renewcommand{\Pr}[1]{\mathrm{Pr}\left\{#1\right\}}
\newcommand{\expect}[2]{\mathsf{E}_{#1}\left\{#2\right\}}

\newcommand{\setU}{\mathcal{U}}

\newcommand{\PEP}{p}
\newcommand{\PEPt}{\bar{p}}
\newcommand{\Prss}{\rho}

\newcommand{\setV}{\mathcal{V}}
\newcommand{\setC}{\mathcal{C}}
\newcommand{\setE}{\mathcal{E}}
\newcommand{\setS}{\mathcal{S}}
\newcommand{\setG}{\mathcal{G}}
\newcommand{\setGt}{\tilde{\mathcal{G}}}
\newcommand{\maxd}{q}

\begin{document}

\begin{DIFnomarkup}

\title{Error Floor Analysis of Coded Slotted ALOHA over Packet Erasure Channels}

\author{
%Author 1, Author 2, Author 3, Author 4
Mikhail~Ivanov, Fredrik Br\"{a}nnstr\"{o}m,~\IEEEmembership{Member,~IEEE}, Alexandre Graell i Amat,~\IEEEmembership{Senior Member,~IEEE},\\Petar Popovski,~\IEEEmembership{Senior Member,~IEEE}
\thanks{This research was supported by the Swedish Research Council, Sweden, under Grant No. 2011-5950, the Ericsson's Research Foundation, Sweden, and Chalmers Antenna Systems Excellence Center in the project `Antenna Systems for V2X Communication'. %The calculations were performed on resources provided by the Swedish National Infrastructure for Computing (SNIC) at C3SE.
}
\thanks{M. Ivanov, F. Br\"{a}nnstr\"{o}m, and A. Graell i Amat are with the Department~of Signals and Systems, Chalmers University of Technology, SE-41296 Gothenburg, Sweden (e-mail: \{mikhail.ivanov, fredrik.brannstrom, alexandre.graell\}@chalmers.se).}
\thanks{Petar Popovski is with the Department of Electronic Systems, Aalborg University, 9220 Aalborg, Denmark (e-mail: petarp@es.aau.dk).}
}

\maketitle

\end{DIFnomarkup}

\begin{abstract}
We present a framework for the analysis of the error floor of \gls{CSA} for finite frame lengths over the packet erasure channel. The error floor is caused by stopping sets in the corresponding bipartite graph, whose enumeration is, in general, not a trivial problem. We therefore identify the most dominant stopping sets for the distributions of practical interest. The derived analytical expressions allow us to accurately predict the error floor at low to moderate channel loads and characterize the unequal error protection inherent in \gls{CSA}.
\end{abstract}

% Not needed for conference
%\begin{keywords} 
%	APSK constellation, nonlinear phase
%	noise, optical Kerr-effect, self-phase modulation. 
%\end{keywords}
% Redefine all acronyms that have been defined in the introduction
\glsresetall

\section{Introduction}\label{sec:intro}
\Gls{CSA} has recently been proposed as an \emph{uncoordinated} multiple access technique that can provide large throughputs close to those of coordinated schemes~\cite{Paolini11,Stefanovic13}. The need for \gls{CSA} arises  in scenarios where high throughput is required and coordinated techniques cannot be used due to various reasons, such as random number of devices in machine-to-machine communication. In this work, we assume a scenario where, besides throughput considerations, reliability plays an important role. We therefore focus on the probability of communication failure as our main figure of merit~\cite{Madueno14}.

Different versions of \gls{CSA} have been proposed (see~\cite{Paolini14} for the most recent review). All of them share a slotted structure borrowed from the original slotted ALOHA~\cite{Roberts75} and the use of successive interference cancellation. The contending users introduce redundancy by encoding their messages into multiple packets, which are transmitted to the \gls{BS} in randomly chosen slots. The \gls{BS} buffers the received signal, decodes the packets from the slots with no collision and attempts to reconstruct the packets in collision exploiting the introduced redundancy. A packet that is reconstructed is subtracted from the buffered signal and the \gls{BS} proceeds with another round of decoding.

The analysis of \gls{CSA} is usually done assuming that a packet in a collision-free slot can be reliably decoded and the interference caused by a packet can be ideally subtracted if the packet is known. Under these assumptions, the system can be viewed as a graph-based code operating over a \gls{BEC}. %Although very idealized, this predicts well the performance of the actual system over the \gls{AWGN} channel for certain physical layer implementations~\cite{Liva11}. 
Most papers on \gls{CSA} consider error-free transmission. Here, we consider transmission over the \gls{PEC}~\cite{Fabregas06}. The \gls{PEC} can be used to model packets that are erased due to a deep fade for a block fading channel in the large \gls{SNR} regime. It can also be used to model a network with random connectivity.

%\mi{ However, these assumptions do not capture some wireless channel effects such as fading. A simple way to deal with fading was suggested in~\cite{Stefanovic13} which, however, requires extra signalling from the \gls{BS} to estimate the channel. Other approaches as in~\cite{Herrero14} and~\cite{Stefanovic14} are much more involved and cannot rely on the \gls{BEC} model. In this paper, we consider a simplified fading channel, i.e., a \gls{PEC}~\cite{Fabregas06} which can easily be treated within the \gls{BEC} framework.}

The \gls{BEC} model allows us to use \gls{DE} to predict the asymptotic performance of the system when the frame length tends to infinity. The typical performance exhibits a threshold behavior, i.e., all users are reliably resolved if the number of users does not exceed a certain threshold. Most of the work on \gls{CSA} focuses on optimizing the threshold. However, the performance in the finite frame length regime is more relevant in practice. Similarly to \gls{LDPC} codes, a finite frame length gives rise to an error floor due to stopping sets present in the graph~\cite{Di02}. For regular \gls{CSA}, some simple approximations to predict the error floor were proposed in~\cite{Herrero14}. In this paper, we derive analytical expressions to predict the error floor for irregular \gls{CSA}~\cite{Liva11} over the \gls{PEC}. The derived expressions also allow us to predict the \gls{UEP} inherent in irregular \gls{CSA}. Moreover, we show that the asymptotic analysis using \gls{DE} fails for systems operating over the \gls{PEC}.

\section{System Model}\label{sec:syst_model}

\begin{figure}
	\centering
	\subfloat[Time representaion of CSA.]{\label{fig:system_model_a}
	\includegraphics{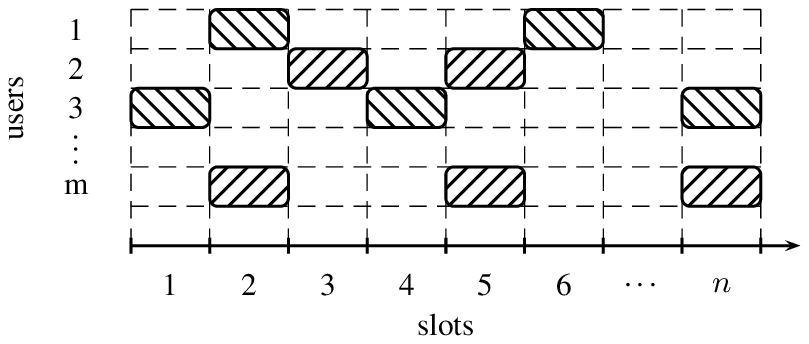}
	}
	
	\subfloat[Graph representation of CSA.]{\label{fig:system_model_b}
	\includegraphics{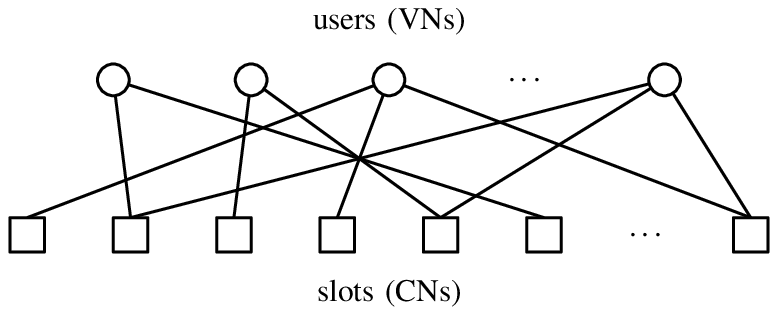}
	}
	\caption{System model.}
	\vspace{-0.4cm}
	\label{fig:system_model}
\end{figure}

We consider $m$ users that transmit to the~\gls{BS} over a shared medium. The communication takes place during a contention period called \emph{frame}, consisting of $n$ slots of equal duration. 
Using a properly designed physical layer\footnote{A particular implementation of the physical layer is not important in the context of this paper. However, it greatly affects the system performance over more realistic channels.}, each user maps its message to a physical layer packet and then repeats it $l$ times ($l$ is a random number chosen based on a predefined distribution) in randomly chosen slots, as shown in~\figref{fig:system_model}\subref{fig:system_model_a}. This setup can be viewed as repetition coding and such a user is called a degree-$l$ user. Every packet contains pointers to its copies, so that, once a packet is successfully decoded, full information about the location of the copies is available.

The received signal at the \gls{BS} in the $i$th slot is
\begin{equation}
	y_i = \sum_{j \in \setU_i} h_{i,j} a_{j},
\end{equation}
where $a_j$ is a packet of the $j$th user, $h_{i,j}$ is the channel coefficient and $\setU_i \subset \{1,\dots, m\}$ is the set of users that transmit in the $i$th slot. We assume the channel coefficients to be independent across users and slots and identically distributed such that $\Pr{|h_{i,j}| = 0} = \epsilon$ and $\Pr{|h_{i,j}| > 0} = 1- \epsilon$, i.e., $\epsilon$ is the probability of a packet erasure. We refer to such a channel as a \gls{PEC}. If the number of nonzero coefficients $h_{i,j} \neq 0$ for $j \in \setU_i$ is one, the $i$th slot is called a \emph{singleton} slot. Otherwise, we say that a collision occurs in the $i$th slot.

At first the \gls{BS} decodes the packets in singleton slots and obtains the location of their copies. Using data-aided methods, the channel coefficients corresponding to the copies are then estimated. After subtracting the interference caused by the identified copies, decoding proceeds until no further singleton slots are found. The \gls{PEC} also models the capture effect since a packet in the $i$th slot may be decoded even if $|\setU_i| > 1$.

The system can be analyzed using the theory of codes on graphs over the \gls{BEC}. Each user corresponds to a \gls{VN} and represents a repetition code, whereas slots  correspond to \glspl{CN} and can be seen as single parity-check codes. In the following, users and \glspl{VN} are used interchangeably. An edge connects the $j$th \gls{VN} to the $i$th \gls{CN} if $j \in \setU_i$ and $h_{i,j} \neq 0$. For the example in~\figref{fig:system_model}\subref{fig:system_model_a}, the corresponding bipartite graph is shown in~\figref{fig:system_model}\subref{fig:system_model_b}. A bipartite graph is defined as $\setG =\{\setV, \setC, \setE\}$, where $\setV$, $\setC$, and $\setE$ represent the sets of \glspl{VN}, \glspl{CN}, and edges, respectively. The performance of the system greatly depends on the distribution that users use to choose the degree $l$ or, using graph terminology, on the \gls{VN} degree distribution
\begin{equation}\label{eq:distr_orig}
	\tilde{\lambda}(x) = \sum_{l = 0}^{\maxd}\tilde{\lambda}_{l}x^{l},
\end{equation}
where $x$ is a dummy variable, $\tilde{\lambda}_l$ is the probability of choosing degree $l$, and $\maxd$ is the maximum degree, which is often bounded due to implementation constraints, for instance, by eight in~\cite{Liva11}. We define a vector representation of~\eqref{eq:distr_orig} as $\bm{\tilde{\lambda}} = [\tilde{\lambda}_0, \dots, \tilde{\lambda}_q]$. For a graph $\setGt$ generated using $\bm{\tilde{\lambda}}$, we define the graph profile as the vector $\bm{v}(\setGt) = [ v_0(\setGt), v_1(\setGt), \dots, v_q(\setGt)]$, where $v_l(\setGt)$ is the number of degree-$l$ \glspl{VN} in $\setGt$. The profile of a random graph $\setGt$ has a distribution $\Pr{\bm{v}(\setGt) = \bm{u}}= p_{\mathsf{mn}}(\bm{u}, \bm{\tilde{\lambda}}, m)$, where 
\begin{equation}\label{eq:multinomial}
 p_{\mathsf{mn}}(\bm{u}, \bm{\tilde{\lambda}}, m) = \begin{cases} m! \prod_{l = 0}^{\maxd}\frac{\tilde{\lambda}_{l}^{u_{l}}}{u_l!} & \text{if } \| \bm{u} \|_1 = m, \\
 0 & \text{otherwise}
 \end{cases}
\end{equation}
is the multinomial distribution and $\|\cdot \|_1$ is the $\ell_1$ norm.

The key performance parameters are defined as follows. The channel load $g = m/n$ shows how ``busy'' the medium is. The average number of users that successfully transmit their message, termed \emph{resolved users}, is denoted by $r$. The throughput $t = r/n$ shows how efficiently the frame is used. In this paper, we focus on the average \gls{PLR} $\PEPt = (m-r)/m = 1 - t/g$, which is the fraction of unresolved users, i.e., the users whose messages are not successfully decoded by the \gls{BS}.

\section{Performance Analysis}

Let a user repeat its packet $l$ times. Each copy is erased with probability $\epsilon$. Hence, the \gls{BS} receives $k \in \{0, \dots, l\}$ packets with probability $\binom{l}{k} \epsilon^{l-k} (1-\epsilon)^{k}$. Averaging over the \gls{VN} degree distribution $\tilde{\lambda}(x)$ leads to the \emph{induced} \gls{VN} degree distribution $\lambda(x)$ observed by the \gls{BS}\footnote{We use tilde to denote quantities related to the original distribution; the analogous quantities without tilde correspond to the induced distribution.}
\begin{equation}\label{eq:induced_distribution_pre}
	\lambda(x) = \sum_{l = 0}^{\maxd}\tilde{\lambda}_{l} \sum_{k = 0}^{l} \binom{l}{k} \epsilon^{l-k} (1-\epsilon)^{k}x^{k},
\end{equation}
%Combining the terms of the same degree, \eqref{eq:induced_distribution_pre}
which can be written similarly  to~\eqref{eq:distr_orig}, where
\begin{equation}\label{eq:lambda}
	\lambda_{l} = \sum_{k = l}^{\maxd}\binom{k}{l} \epsilon^{k-l} (1-\epsilon)^{l}\tilde{\lambda}_{k}
\end{equation}
is the fraction of users of degree $l$ as observed by the \gls{BS}.
%For example, the distribution~\cite{Liva11}
%\begin{equation}\label{eq:dist_example}
%\lambda(x) = 0.25 x^2 + 0.6 x^3 + 0.15x^8,
%\end{equation}
%will be turned into the distribution
%\begin{align}
%\lambda'(x) =&  2.4\cdot 10^{-4}  + 1.6\cdot 10^{-2}x^1 + 2.9\cdot 10^{-1}x^2\nonumber\\
%		& + 5.5\cdot 10^{-1}x^3  +  7.5\cdot 10^{-6}x^4  + 2.0\cdot 10^{-4}x^5\nonumber\\
%		 & +  3.1\cdot 10^{-3}x^6  + 2.9\cdot 10^{-2}x^7 + 1.2\cdot 10^{-1}x^8
%\end{align}
%by the \gls{PEC} with $p = 0.03$.
The \gls{PEC} can thus be analyzed by considering  $\lambda(x)$ over a standard collision channel~\cite{Liva11}. The main peculiarity of the induced distribution, not considered in the standard \gls{CSA} analysis, is that it contains all degrees up to $\maxd$, including zero and one. This implies that the \gls{PLR} exhibits an error floor, which is lowerbounded by $\lambda_0 = \tilde{\lambda}(\epsilon) >0$ for $\epsilon > 0$. Hence, the threshold, i.e., the channel load below which the \gls{PLR} is zero when $n \rightarrow \infty$, predicted by \gls{DE} is zero over a \gls{PEC} with $\epsilon > 0$ for any distribution $\tilde{\lambda}(x)$. Moreover, \gls{DE} devised in~\cite{Liva11} does not correspond to the decoding algorithm for distributions with $\lambda_1 \neq 0$. In the following, we focus on the induced distribution $\lambda(x)$ and the induced graph $\setG = \{\setV, \setC, \setE\}$.

By construction \gls{CSA} features \gls{UEP}, similarly to the \gls{UEP} in coding theory, since users with different degrees are protected differently\footnote{We note, however, that in a series of contention rounds, each user will have the average performance if the degree is chosen randomly each time.}. The inherent \gls{UEP} is illustrated by dashed lines in~\figref{fig:sim_results}\subref{fig:num_res_e0} for $\tilde{\lambda}(x) = 0.25 x^2 + 0.6 x^3 + 0.15 x^8$ and $\epsilon = 0$. As clearly seen from the figure, the higher the degree, the better the \gls{PLR} performance. We remark that the performance of a degree-$l$ user depends on the entire distribution and not only on its degree. To characterize the performance of users of different degrees, we define the \gls{PLR} for a degree-$l$ user as observed by the \gls{BS} as
\begin{equation}\label{eq:pep_l}
\PEP_l = \frac{\bar{w}_l}{\bar{m}_l} = \frac{\bar{w}_l}{ m \lambda_l},
\end{equation}
where $\bar{m}_l$ and $\bar{w}_l$ are the average number of all and unresolved degree-$l$ users, respectively. For degree-$0$ users, $p_0=1$. The average \gls{PLR} is
\begin{equation}\label{eq:plr_tot}
\PEPt = \frac{1}{m}\sum_{l = 0}^{q}{\bar{w}_l} = \sum_{l = 0}^{\maxd}{\lambda_l \PEP_l}.
\end{equation}

%Obviously, users that never transmit ($l = 0$) will never get their packet across, whereas users that transmit many times have a high chance to be resolved. We call this property \gls{UEP} similarly to the \gls{UEP} in coding theory.\footnote{We note, however, that in a series of contention rounds, each user will have the average performance if the degree is chosen randomly each time.} This property can be exploited, for instance, to assign different protection levels to users with different ``priorities''. However, when a user starts to choose the degree based on its priority, the performance of the whole system and the performance of each user can be greatly affected. Exploiting the \gls{UEP} in an uncoordinated fashion for \gls{CSA} is left for future investigation. An example of a typical \gls{PLR} performance for users of different degrees is shown in~\figref{fig:typical_perf} with dashed lines with markers.

%\begin{figure}
%	\includegraphics[]{../pstricks/uniform_distribution/uniform}
%	\caption{The PLR performance for the uniform distribution $\tilde{\lambda}(x) = 0.2x +  0.2x^2 + 0.2 x^3 + 0.2x^4 + 0.2x^5$ and $n=200$ slots. Dashed lines with markers show simulation results, solid lines show analytical error floor approximations.}
%	\label{fig:typical_perf}
%\end{figure}

\begin{figure}
\vspace{-0.34cm}
	\centering
	\subfloat[$\setS_1$.]{\label{fig:type1}
	\includegraphics{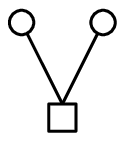}
	}
	\subfloat[$\setS_2$.]{\label{fig:type2}
	\includegraphics{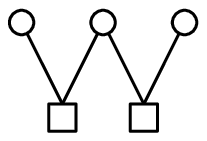}
	}
	\subfloat[$\setS_3$.]{
	\includegraphics{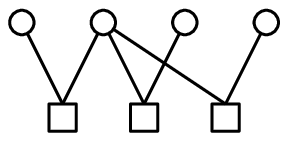}
	}
	\subfloat[$\setS_4$.]{\label{fig:type4}
	\includegraphics{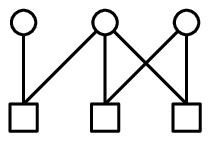}
	}
	
	\subfloat[$\setS_5$.]{\label{fig:two_two}
	\includegraphics{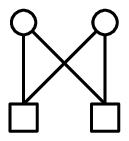}
	}
	\subfloat[$\setS_6$.]{
	\includegraphics{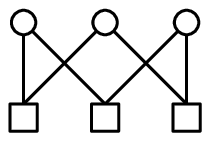}
	}	
	\subfloat[$\setS_7$.]{
	\includegraphics{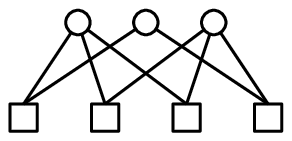}
	}
	%\qquad
	\subfloat[$\setS_8$.]{\label{fig:type8}
	\includegraphics{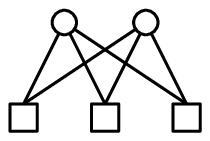}
	}
	\caption{Stopping sets.}
	\vspace{-0.4cm}
	\label{fig:error_events}
\end{figure}

Since packet erasures are accounted for in the induced distribution, the only source of errors in the considered model is harmful structures in the graph $\mathcal{G}$. When, e.g., two degree-$2$ users transmit in the same slots (see~\figref{fig:error_events}\subref{fig:two_two}), the \gls{BS} will not be able to resolve them. Such harmful structures are commonly referred to as \emph{stopping sets}. A subset of  \glspl{VN} of non-zero degrees $\setS \subset \setV$ forms a stopping set if all neighbors of $\setS$ are connected to $\setS$ at least twice~\cite{Di02}. $\setS$ induces a graph, which, with a slight abuse of notation, is also referred to as stopping set and denoted by $\setS$. Therefore, $\setS$ can be described by its profile $\bm{v}(\setS)$.
For $\setS_2$ in~\figref{fig:error_events}\subref{fig:type2},  the graph profile is $\bm{v}(\setS_2) = [0, 2, 1,0,\dots, 0]$, where the number of zeros in the end depends on $q$.

Stopping sets are referred to as ``loops'' in~\cite{Herrero14}. However, stopping sets do not form a loop if degree-$1$ users are present, as in~\figref{fig:error_events}\subref{fig:type1}--\subref{fig:type4}. On the other hand, if there are no degree-$1$ users, all stopping sets create cycles (or loops) in the graph, as in~\figref{fig:error_events}\subref{fig:two_two}--\subref{fig:type8}. Length-$4$ cycles are the shortest cycles  possible  and are always avoided in properly designed \gls{LDPC} codes. Due to the randomly generated graph, however, length-$4$ cycles are intrinsic to \gls{CSA}. %We remark that not all length-$4$ cycles are harmful for ~\gls{CSA}, that is, if a degree-3 user collides with the degree-2 user in two slots, such a situation can be resolved. 
%Avoiding short cycles in a randomly generated graph is left for future work.

We denote the probability of a stopping set $\setS$ to occur by $\Prss(\setS)$. Let $\mathcal{A}$ be the set of all possible stopping sets. Using a union bound argument, the number of unresolved degree-$l$ users can be upperbounded as
\begin{equation}\label{eq:error_number}
	\bar{w}_l \le \sum_{\setS \in \mathcal{A}}{v_{l}(\setS) \Prss(\setS)}.
\end{equation}

The probability $\Prss(\setS)$ is in general difficult to evaluate. In the following, we give an example to show how it can be approximated. Consider $\setS_5$ in~\figref{fig:error_events}\subref{fig:two_two} and assume a particular realization of $\setG$ so that $v_2(\setG) \ge 2$. Numerical results show that for low channel loads, the probability to observe exactly one stopping set $\setS_5$ is at least one order of magnitude larger than the probability of any other possible stopping set formed by degree-$2$ users (including multiple $\setS_5$) to occur. We therefore assume that only one stopping set $\setS_5$ occurs. This also hints that $\setS_5$ is the most dominant among all stopping sets containing only degree-$2$ users. Users of other degrees may transmit in the same slot as the users in $\setS_5$. This may create a new larger stopping set, or these users may be successfully resolved and their interference subtracted. However, the users in $\setS_5$ will remain unresolved. Each of the degree-$2$ users has $\gamma = \binom{n}{2}$ possible combinations of slots for transmission. Given that the number of slots is large enough for the realization of $\setG$ ($\gamma \ge v_2(\setG)$), we write
\begin{equation}\label{eq:rho_1_ex}
	\rho(\setS_5, \setG) \approx \alpha(\setS_5, \setG) \beta(\setS_5, \setG),
\end{equation}
where
\begin{equation}\label{eq:multipl_s5}
	\alpha(\setS_5, \setG) = \binom{v_2(\setG)}{2}
\end{equation}
and
\begin{equation*}
\beta(\setS_5, \setG) =\frac{1}{\gamma^{v_2(\setG)}}\!\!\prod_{k = 0}^{v_2(\setG)-2}\!\! (\gamma -k) =  \frac{\gamma!}{(\gamma-v_2(\setG) + 1)! \gamma^{v_2(\setG)}},
\end{equation*}
where the right-hand side of~\eqref{eq:rho_1_ex} is the probability of one stopping set $\setS_5$ to occur and is split into a product of two variables for future~use.

In the following, we remove the dependency on $\setG$ from~\eqref{eq:rho_1_ex} in order to obtain $\rho(\setS_5)$. To do so, we simplify $\beta(\setS_5, \setG)$ assuming that $\gamma \gg v_2(\setG)$. By letting $\gamma$ grow large, we obtain
$\beta(\setS_5, \setG) \approx \beta(\setS_5) = \gamma^{-1}$. The same result can be obtained setting $v_2(\setG)$ to the minimal possible value, i.e., two. To find $\rho(\setS_5)$, $\alpha(\setS_5, \setG)$ in~\eqref{eq:multipl_s5} needs to be averaged over all $\setG$, i.e.,
%\begin{equation}\label{eq:multiplicity}
	$\alpha(\setS_5) = \expect{\setG}{\alpha(\setS_5, \setG)}$,
%\end{equation}
where $\expect{\setG}{\cdot}$ stands for the expectation over $\setG$.

For a generic $\setS$, \eqref{eq:multipl_s5} generalizes to
\begin{equation}\label{eq:multi_general}
\alpha(\setS, \setG) = \prod_{l = 1}^{\maxd} \binom{v_l(\setG)}{v_l(\setS)}.
\end{equation}
Using~\eqref{eq:multi_general} and the definition of the multinomial distribution in~\eqref{eq:multinomial}, $\alpha(\setS)= \expect{\setG}{\alpha(\setS, \setG)}$ can be calculated as
\begin{equation}\label{eq:multi_final}
%	\alpha(\setS) =  \frac{m!}{(m - v_{\Sigma}(\setS))!} \prod_{l = 1}^{\maxd}\frac{\lambda_{l}^{v_{l}(\setS)}}{v_l(\setS)!}.
\alpha(\setS) =  \binom{m}{\|\bm{v}(\setS)\|_1}p_{\mathsf{mn}}(\bm{v}(\setS), \bm{\lambda}, \|\bm{v}(\setS)\|_1).
\end{equation}

%\begin{figure*}[!t]
%\normalsize
%\begin{multline}
%\label{eq:long_eq}
%\alpha =\frac{m!}{\prod_{i = 1}^{\maxd} v_i!} 
%\sum_{l'_1}\mkern-45mu\sum_{\substack{l'_2\\  l'_i \in \{0, \dots, m-v_{\Sigma}\} , i = 1, \dots, \maxd\\ \sum_{i=0}^{\maxd} l'_i \le m-v_{\Sigma}}}\mkern-50mu \cdots \sum_{l'_{\maxd}}
% \frac{\prod_{i = 1}^{\maxd}\lambda_{i}^{l'_{i} + v_i} \lambda_0^{(m- v_{\Sigma} - \sum_{i = 1}^{\maxd} l'_i)}}{(m- v_{\Sigma} - \sum_{i = 1}^{\maxd} l'_i)!\prod_{i = 1}^{\maxd} l'_i!}
%\\=m!\prod_{i = 1}^{\maxd} \frac{\lambda_i^{v_i}}{ v_i!} 
%\sum_{l'_1}\mkern-45mu\sum_{\substack{l'_2\\  l'_i \in \{0, \dots, m-v_{\Sigma}\} , i = 1, \dots, \maxd\\ \sum_{i=0}^{\maxd} l'_i \le m-v_{\Sigma}}}\mkern-50mu \cdots \sum_{l'_{\maxd}}
% \frac{\prod_{i = 1}^{\maxd}\lambda_{i}^{l'_{i}} \lambda_0^{(m- v_{\Sigma} - \sum_{i = 1}^{\maxd} l'_i)}}{(m- v_{\Sigma} - \sum_{i = 1}^{\maxd} l'_i)!\prod_{i = 1}^{\maxd} l'_i!}.
%\end{multline}
%\hrulefill
%\vspace*{4pt}
%\end{figure*}

We apply the reasoning above to all stopping sets and write
\begin{equation}\label{eq:approximation}
\Prss(\setS) \approx \alpha(\setS) \beta(\setS),
\end{equation}
where $\beta(\setS)$ is the probability of the stopping set $\setS$ to occur in a graph with  profile $\bm{v}(\setS)$, and $\alpha(\setS)$ is the number of combinations to choose $v_l(\setS)$ users out of $v_l(\setG)$ users present in the graph $\setG$ for all degrees, averaged over $\setG$. The numerical results presented later on justify the approximation in~\eqref{eq:approximation}.

Identifying all stopping sets and calculating the corresponding $\beta(\setS)$ in a systematic way is not possible in general. In practice, distributions with large fractions of low-degree \glspl{VN} are most commonly used since they achieve high thresholds. For instance, the soliton distribution~\cite{Narayanan12}, which asymptotically provides throughput arbitrarily close to one, has $\tilde{\lambda}_2 = 0.5$ and  $\tilde{\lambda}_3 = 0.17$.  If we constrain ourselves to the family of such distributions, i.e., distributions with large fractions of degree-$2$ and degree-$3$ users, identifying the most dominant stopping sets becomes possible. By running extensive simulations for different distributions from the aforementioned family of distributions for $g = 0.5$, we determined the stopping sets that contribute the most to the error floor by estimating their relative frequencies. These stopping sets are shown in~\figref{fig:error_events}, with
\begin{align}
	&\beta(\setS_1) = \frac{1}{n}, \beta(\setS_2) = \frac{2}{n^2}, \beta(\setS_3) = \frac{6}{n^3},\beta(\setS_4) = \frac{6}{(n-1)n^2},\nonumber\\ 
	&\beta(\setS_5) = \frac{2}{(n-1)n},
	\beta(\setS_6) = \! \frac{4(n-3)}{(n\!-\!2)n^3}, \label{eq:pn}\\
	&\beta(\setS_7) =  \frac{36(n-3)}{(n-2)(n-1)n^3},
	\beta(\setS_8) =  \frac{6}{(n-2)(n-1)n}.\nonumber
\end{align}
%\mi{For a particular distribution, some of the stopping sets may not be relevant.} 
Note that constraining the set of the considered stopping sets turns the upper bound  in~\eqref{eq:error_number} into an approximation. The \gls{PLR} of a degree-$l$ user can thus be approximated as
\begin{equation}\label{eq:final_aprx}
	p_l \approx \frac{1}{m\lambda_l}\sum_{\setS \in \mathcal{A}_8}{v_{l}(\setS) \alpha(\setS) \beta(\setS)},
\end{equation}
where $\mathcal{A}_8$ is the set of the eight stopping sets in~\figref{fig:error_events} with the corresponding $\alpha(\setS)$ in~\eqref{eq:multi_final} and $\beta(\setS)$ in~\eqref{eq:pn}.

The analysis above is presented for the induced degree distribution, i.e., $p_l$ in~\eqref{eq:pep_l} is the \gls{PLR} for a user of degree $l$ as observed by the \gls{BS}. The \gls{PLR} for a degree-$l$ user (from the user's perspective) can be found as
\begin{equation}\label{eq:orig_err}
	\tilde{p}_{l} = \sum_{k = 0}^{l} p_k \binom{l}{k} \epsilon^{l-k}(1-\epsilon)^k
\end{equation}
if $\tilde{\lambda}_l \neq 0$ and zero otherwise. When $\epsilon = 0$, it is easy to see from~\eqref{eq:lambda} that $\tilde{\lambda}_l = \lambda_l$ and from~\eqref{eq:orig_err} that $\tilde{p}_l = p_l$ for all $l = {0, \dots, q}$. We remark that calculating~\eqref{eq:orig_err} is not required if we are interested in the average \gls{PLR}, i.e., it can be calculated based on $p_l$ and $\lambda(x)$ as in~\eqref{eq:plr_tot}.

%\subsection{Large Frame Length}
%
%\mi{Show how to calculate $\PEP_1$.}
%\begin{equation}
%	\PEP_1 \approx \lambda_1 g + 2 \lambda_1 \lambda_2 g^2 + 3 \lambda_1^2 \lambda_3 g^3.
%\end{equation}
%Analogously, we can calculate the other probabilities
%\begin{align}
%	 \PEP_2 & \approx \lambda_1^2 g^2,\\
%	 \PEP_3 & \approx \lambda_1^3 g^3.
%\end{align}
%
%The error floor can then be written as
%\begin{equation}
%	\PEPt \approx \lambda_0 +  \lambda_1^2 g + 3\lambda_1^2 \lambda_2 g^2 + 4\lambda_1^3 \lambda_3 g^3.
%\end{equation}
%which means that the error floor exists if $\lambda_1 \neq 0$. This also implies that the threshold for such distributions is zero. On the other had, when $\lambda_1 = 0$, the error floor is zero.
%
%\mi{Maybe put a figure here showing how the error floor depends on $n$ for 2-3 distributions? We can also identify the limiting factor, i.e., the error floor due to erasures or due to small $n$.}
\section{Numerical Results}

\begin{figure}
\centering
\vspace{-0.34cm}
	\subfloat[$\epsilon = 0$.]{
		\includegraphics[]{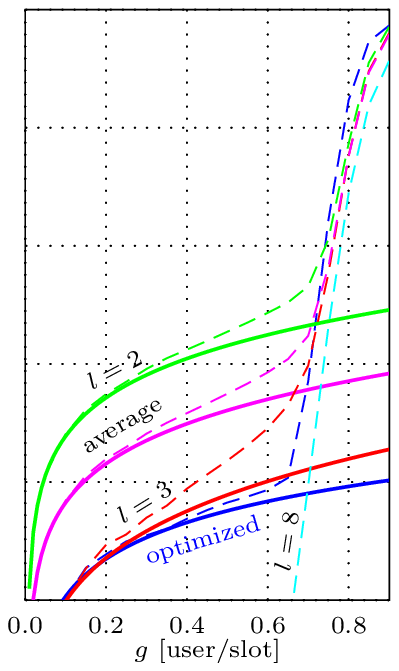}\label{fig:num_res_e0}
	}
	\subfloat[$\epsilon = 0.03$.]{
		\includegraphics[]{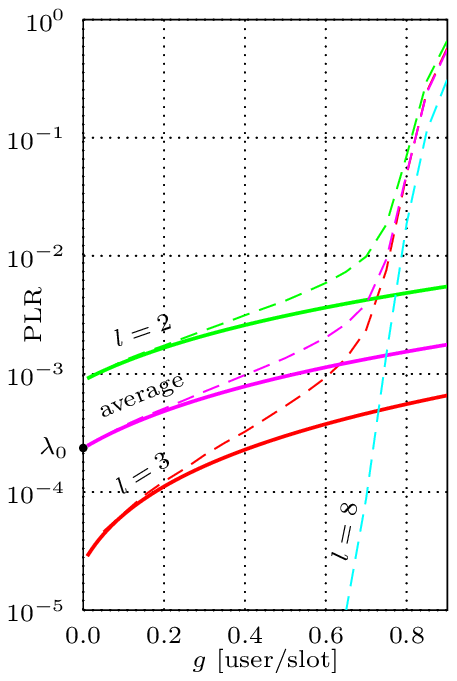}\label{fig:num_res_e003}
	}
	\caption{The \gls{PLR} performance for $n =200$. Dashed lines show simulation results and solid lines show analytical error floor approximations~\eqref{eq:orig_err}.}
	 \vspace{-0.4cm}
	\label{fig:sim_results}
\end{figure}

In~\figref{fig:sim_results}, we plot the \gls{PLR} for a degree-$l$ user (from the user's perspective) for $\tilde{\lambda}(x) = 0.25 x^2 + 0.6x^3 +0.15x^8$ and two different values $\epsilon$. This distribution was suggested in~\cite{Liva11} as a distribution that provides low error floor. The dashed lines show the simulation results and the solid lines show the proposed analytical PLR expression~\eqref{eq:orig_err} based on the approximation~\eqref{eq:final_aprx}. The analytical \gls{PLR} predictions demonstrate good agreement with the simulation results for low to moderate channel loads. This justifies the approximation in~\eqref{eq:approximation} and the use of the stopping sets in~\figref{fig:error_events}. Considering other stopping sets can improve the \gls{PLR} prediction for higher channel loads and would make it possible to predict the performance of users of higher degrees. However, as~\figref{fig:sim_results} suggests, the contribution of these users to the average \gls{PLR} is negligible, especially in the error floor region. We also remark that the accuracy of the approximations improves when the frame length increases. In~\figref{fig:sim_results}\subref{fig:num_res_e003}, we also show the value of $\lambda_0 = \tilde{\lambda}(0.03) = 2.4 \times 10^{-4}$ to demonstrate the lower bound on the average \gls{PLR}.

The proposed \gls{PLR} approximation can also be used for the optimization of the degree distribution. As an example, we performed the optimization for $n = 200$ and $\epsilon = 0$ using a linear combination of the threshold provided by \gls{DE} and the analytical prediction of the average \gls{PLR} as the objective function. One of the obtained distributions is $\tilde{\lambda}(x) = 0.87x^3 + 0.13x^8$ and its average \gls{PLR} performance is shown with blue lines in~\figref{fig:sim_results}\subref{fig:num_res_e0}. This distribution has considerably lower error floor without significant degradation of the performance in the waterfall region compared to the distribution from~\cite{Liva11}.

\section{Conclusions}
In this letter, we proposed analytical approximations of the error floor of \gls{CSA} for finite frame length over the \gls{PEC}. These approximations show good agreement with simulation results for most distributions of practical interest. The approximations can be used to optimize the degree distribution for a given frame length and packet erasure probability.

\bibliographystyle{IEEEtran}
%\bibliography{../bibliography/MyBibliography}


\end{document}